\documentstyle[aps,preprint,epsfig]{revtex}
\tightenlines
\begin{document}
\draft
\title{Phase separation of incompressible binary fluids with 
Lattice Boltzmann Methods}
\author{ Aiguo Xu$^{1}$, G.  Gonnella$^{1,2}$ and  A. Lamura$^{3}$}
\address{
$^{1}$ Istituto Nazionale per la  Fisica della Materia, Unit\`a di Bari,
{\rm and} Dipartimento di Fisica, Universit\`a di Bari, {\rm and}
TIRES, Center of Innovative Technologies for Signal Detection
and Processing,
 via Amendola 173, 70126 Bari, Italy\\
$^{2}$ INFN, Sezione di Bari,  via Amendola 173, 70126 Bari\\
$^{3}$ Istituto Applicazioni Calcolo, CNR, Sezione di Bari,\\
Via Amendola 122/I, 70126 Bari, Italy }
\date{\today}
\maketitle
\begin{abstract}

We introduce new versions of 
lattice Boltzmann methods (LBM) for incompressible binary mixtures
where fluctuations of total density are inhibited. 
As a test for the improved algorithms we consider the  problem of
phase separation of two-dimensional binary mixtures 
quenched from a disordered state into the coexistence region.
We find that the stability properties of LBM
are greatly improved. The control of density fluctuations
and the possibility of running longer simulations
allow a more precise evaluation  of the growth exponent. 

\end{abstract}
\vskip 0.5cm
\pacs{KEY WORDS: Lattice Boltzmann methods; binary mixtures; phase
separation.}

\section{Introduction}

Lattice Boltzmann methods (LBM) are computational schemes for
solving the macroscopic equations of fluid systems \cite{Chen98,succi}.
In the case of binary  mixtures they have been largely 
used to study  the dynamics of  systems 
described by  the Navier-Stokes and convection-diffusion equations \cite{Yeomans}. 
The advantages 
of LBM compared with standard  methods for partial differential equations
are particularly relevant in problems involving complex boundaries such as
multiphase flow in porous media \cite{olson,Chen98,succi}. 
The large number of applications of LBM in
studies on  complex fluids \cite{gonnella,gompper,lamepl}, 
polymer mixtures \cite{Malevanets99}, liquid crystals \cite{yeom},
etc. demonstrates the versatility of the method.
In studies of spinodal decomposition of binary mixtures
a version of LBM based on a free-energy approach,
where the equilibrium distributions can be consistently chosen 
with the thermodynamics of the system, has been largely used \cite{Orlandini,Swift,catesjfm}.

The stability properties of LBM are always crucial,
especially in phase separation studies where long lasting 
simulations are needed to establish the growth properties.
Indeed, the average size of domains of the two phases 
generally grows  with power law behavior and one wants to determine
the exponent \cite{Furukawa85,Bray94}.
However, previous studies showed the existence of  a stability problem:
total density fluctuations can suddenly grow in an uncontrolled
way interrupting simulations and
making problematic the determination of the growth exponent \cite{catesjfm}.

In this paper we consider new versions of LBM 
where  the stability  of the algorithm is improved.
We have checked the stability properties by studying 
the problem of phase separation of two-dimensional binary mixtures.
This is a severe test for stability since
density fluctuations are enhanced by the presence of many 
interfaces in the system though they should be negligible in incompressible
fluids. We find that,
with the new algorithms, density fluctuations are largely suppressed and 
a  more reliable determination of the growth exponent
can  be  obtained.

The  incompressible Navier-Stokes equation can be obtained
from lattice Boltzmann equations if density
fluctuations are negligible \cite{Chen98}. 
Compressibility effects, intrinsic to LBM, 
can produce some serious errors in numerical 
simulations of incompressible fluids.    
Due to  these reasons improved versions of LBM
have been  introduced for the case of a simple fluid \cite{zou95,dembo}.
New schemes are based on the 
explicit elimination in the equilibrium distributions of
 higher order terms in the Mach number \cite{heluo} or on the
introduction of  a feedback mechanism \cite{fang}.
With these improvements it has been shown that 
analytical solutions known for particular flows  
are better reproduced than with old LBM versions.

The above methods are here
developed and introduced for the first time 
in the case of binary mixtures.
The algorithms are presented in Section II.
In Section III we study the
problem of phase separation comparing the results
obtained with different LBM versions.
The role  of boundary conditions on the problem of stability
is examined in Section IV. Finally, in  Section V,
we discuss our results and draw our conclusions.

\section{The model}

Our simulations are based on the free-energy approach 
developed by Orlandini {\it et al.} \cite{Orlandini} and Swift {\it et al.} 
\cite{Swift}.
In this scheme a suitable free energy is introduced to control the
equilibrium properties of the system. 

\subsection{The lattice Boltzmann scheme}

The free-energy functional used in the present study is 
\begin{equation}
\begin{cal} F \end{cal}= \int d {\bf r}\left[ \frac{1}{3} n \ln n 
+ \frac{a}{2}\varphi^2
+\frac{b}{4}\varphi^4+\frac{\kappa}{2} (\nabla \varphi)^{2}\right]
\label{fren}
\end{equation}
where $n$ is the total density and $\varphi$ is the density difference
between the two fluids.
The term in $n$ gives rise to a positive background pressure. 
The terms in $\varphi$ in the free-energy density $f(n, \varphi)$ correspond 
to the usual 
Ginzburg-Landau expression typically used in coarse-grained models
of phase separation \cite{Bray94}.
The polynomial terms relate to the bulk
properties of the fluid.  
While the parameter $b$ is always positive,
$a$ allows to distinguish between a disordered ($a>0$)
and a segregated mixture ($a<0$) where two pure phases
with $\varphi = \pm \sqrt{-a/b}$ coexist. 
We will consider quenches into the coexistence region 
with $a<0$ and $b=-a$ so that the equilibrium values for the order 
parameter are $\varphi = \pm 1$.
The gradient term is related to the surface tension.
The other thermodynamic properties of the fluid 
follow from the free energy (\ref{fren}).
The chemical potential difference between the two fluids is given by
$\Delta \mu =  \delta{\cal F}/\delta\varphi=
a \varphi + b \varphi^3 - \kappa \nabla^2 \varphi$.   
The thermodynamic pressure tensor $P_{\alpha\beta}^{th}$ is given by
\begin{eqnarray}
P_{\alpha\beta}^{th}&=&
\Big ( n \; \frac{\delta{\cal F}}{\delta n}
+\varphi \; \frac{\delta{\cal F}}{\delta\varphi} - f(n, \varphi) \Big )
\delta_{\alpha\beta}+\kappa \partial_{\alpha} \varphi 
\partial_{\beta} \varphi
\nonumber \\
&=&\frac{1}{3} n \delta_{\alpha\beta}
+ P_{\alpha\beta}^{chem}
\label{pres}
\end{eqnarray}
where the chemical part is 
$\displaystyle P_{\alpha\beta}^{chem} = \Big (\frac{a}{2} \varphi^2 +
\frac{3}{4} b
\varphi^4 -\kappa \varphi \left( \nabla^2 \varphi \right) -\frac{\kappa}{2}
\left( \nabla \varphi \right)^2 \Big ) \delta_{\alpha\beta} + 
\kappa \partial_{\alpha} \varphi 
\partial_{\beta} \varphi$.

The lattice Boltzmann model is defined on the two-dimensional square lattice
with the following nine velocity vectors:
${\bf e}_{0}=(0,0)$, ${\bf e}_{i} = (\cos [(i-1)\pi/2],
\sin [(i-1)\pi/2 ]) c$,
$i=1,2,3,4$, 
${\bf e}_{i} = (\cos[(i-5)\pi/2+\pi/4],\sin[(i-5)\pi/2+\pi/4]) \sqrt{2} c$,
$i=5,6,7,8$, where $c=\Delta x/\Delta t$, and $\Delta x$ and $\Delta t$ are
the lattice constant and the time step, respectively.
Two sets of
distribution functions $f_{i}({\bf r}, t)$ and $g_{i}({\bf r}, t)$
evolve according to   
a single relaxation-time Boltzmann equation that is discrete in both time and 
space \cite{bhatnagar,chen1}:
\begin{eqnarray}   
f_{i}({\bf r}+{\bf e}_{i}\Delta t, t+\Delta t)-f_{i}({\bf r}, t)&=&   
-\frac{1}{\tau}[f_{i}({\bf r}, t)-f_{i}^{eq}({\bf r}, t)], \label{dist1}\\   
g_{i}({\bf r}+{\bf e}_{i}\Delta t, t+\Delta t)-g_{i}({\bf r}, t)&=&   
-\frac{1}{\tau_{\varphi}}[g_{i}({\bf r}, t)-g_{i}^{eq}({\bf r},   
t)], \label{dist2} 
\end{eqnarray}   
where $\tau$ and  $\tau_{\varphi}$ are independent   
relaxation parameters. 
The distribution functions are related to the total density $n$,  
to the fluid momentum $n {\bf v}$ 
and to the density difference $\varphi$ through
\begin{equation}   
n=\sum_{i}f_{i} , \hspace{1.3cm} n{\bf v}=\sum_{i}f_{i}{\bf e}_{i} ,
\hspace{1.3cm}   
\varphi=\sum_{i}g_{i} .
\label{phys}   
\end{equation}  
These quantities are locally conserved in any collision process and, 
therefore, we
 require that the local equilibrium distribution functions 
$f_{i}^{eq}({\bf r}, t)$ and 
$g_{i}^{eq}({\bf r}, t)$ 
fulfill the Eqs.~(\ref{phys}).
The higher momenta of the local 
equilibrium distribution functions are defined so that the 
continuum equations pertinent to a binary fluid mixture can be obtained 
\cite{Orlandini,Swift}
\begin{equation}   
\sum_{i}f_{i}^{eq}e_{i\alpha}e_{i\beta}=c^2 P_{\alpha\beta}^{th}
+n v_{\alpha} v_{\beta} \;,   
\label{eqn0}
\end{equation}  
\begin{equation}  
 \sum_{i}g_{i}^{eq}e_{i\alpha}=\varphi v_{\alpha} \; ,
\label{eqn} 
\end{equation}
\begin{equation}
\sum_{i}g_{i}^{eq}e_{i\alpha}   
e_{i\beta}=c^2 \Gamma \Delta\mu\delta_{\alpha\beta}+\varphi   
v_{\alpha}v_{\beta} \;.
\label{eqn6}  
\end{equation}   
where $\Gamma$ is a coefficient related to the mobility of the fluid.
The local 
equilibrium distribution functions can be expressed as 
an expansion at the second order in the 
velocity ${\bf v}$ \cite{Orlandini,Swift}. 
\begin{eqnarray}
f_0^{eq}&=& A_0+C_0 v^2 \nonumber\\
f_i^{eq}&=& A_I+B_I v_\alpha e_{i\alpha}+C_I v^2+D_I v_\alpha v_\beta
e_{i\alpha} e_{i\beta}+ G_{I,\alpha\beta}e_{i\alpha} e_{i\beta}
\;\;\;\; i=1,2,3,4 \label{svil1}\\
f_i^{eq}&=& A_{II}+B_{II} v_\alpha e_{i\alpha}+C_{II} v^2
+D_{II} v_\alpha v_\beta
e_{i\alpha} e_{i\beta}+ G_{II,\alpha\beta}e_{i\alpha} e_{i\beta}
\;\;\;\; i=5,6,7,8 \nonumber
\end{eqnarray}
and similarly for the $g_i^{eq}$, $i =0, ..., 8$.
The coefficients in the previous expansions are explicitly written
for convenience to compare the original scheme with 
the alternative ones introduced in the following subsections.
A suitable choice is \cite{aiguo}
\begin{equation}
A_0=\frac{4}{9} n - \frac{5}{6} P_{\alpha\beta}^{chem}\delta_{\alpha\beta}, 
\hspace{0.5cm} A_I=\frac{1}{9} n + \frac{1}{6} P_{\alpha\beta}^{chem}
\delta_{\alpha\beta}, \hspace{0.5cm} 
A_{II}=\frac{A_I}{4}
\label{as}
\end{equation}
\begin{equation}
B_I=\frac{n}{3 c^2}, \hspace{0.5cm} 
B_{II}=\frac{B_I}{4}
\label{bs}
\end{equation}
\begin{equation}
C_0=-\frac{2 n}{3 c^2}, 
\hspace{0.5cm} C_I=-\frac{n}{6 c^2}, \hspace{0.5cm} 
C_{II}=\frac{C_I}{4}
\label{cs}
\end{equation}
\begin{equation}
D_I=\frac{n}{2 c^4}, \hspace{0.5cm} 
D_{II}=\frac{D_I}{4}
\label{ds}
\end{equation}
\begin{equation}
\hspace{-0.2cm} G_{I,\alpha\beta}\!=\!\frac{P_{\alpha \beta}^{th} 
- \frac{1}{2} P_{\sigma \sigma}^{th} \delta_{\alpha \beta}}{2 c^2}, 
\hspace{0.5cm}
 G_{II, \alpha \beta}\!=\!\frac{G_{I,\alpha\beta}}{4}
\label{gs}
\end{equation}
The expansion coefficients for the $g_i^{eq}$ can be obtained
 from the previous ones
with the formal substitutions $n \rightarrow \varphi$ and 
$P_{\alpha \beta}^{th} \rightarrow \Gamma \Delta \mu \delta_{\alpha \beta}$.
The quantities
 $P_{\alpha \beta}^{th}$ and $\Delta \mu$, which appear in the coefficients
of the equilibrium distribution functions, 
can be calculated from (\ref{fren}).

It has been shown in Refs.~\cite{Orlandini,Swift}, 
using a Chapman-Enskog expansion \cite{chapman},
that the above 
described lattice Boltzmann scheme simulates at second order in $\Delta t$
the continuity equation
\begin{equation}
\partial_t n + \partial_{\alpha} (n v_{\alpha}) = 0 + o(\Delta t^2),
\label{cont0}
\end{equation}
and the 
Navier-Stokes equation
\begin{eqnarray}
&& \partial_t (n v_{\beta}) + \partial_{\alpha} (n v_{\alpha} v_{\beta}) =
- \partial_{\beta} p - \partial_{\alpha} c^2 P_{\alpha\beta}^{chem} 
+ \nu \partial_{\gamma} \Big ( n (\partial_{\beta} v_{\gamma}+
\partial_{\gamma} v_{\beta}) \Big) + \nonumber \\
&& +3 \nu \partial_{\alpha} \Big( \partial_t P_{\alpha\beta}^{chem}
- \partial_{\gamma} (n v_{\alpha} v_{\beta} v_{\gamma})
- (v_{\beta} \partial_{\gamma} P_{\alpha\gamma}^{chem}
+ v_{\alpha} \partial_{\gamma} P_{\beta\gamma}^{chem}) \Big) + o(\Delta t^2)
\label{ns0}
\end{eqnarray}
where the isotropic pressure contribution is 
\begin{equation}
p=c_s^2 n ,
\label{presscal}
\end{equation}
$c_s=c/\sqrt{3}$ and $\nu=(\tau - 1/2) c_s^2 \Delta t$ being the speed of sound
and the kinematic viscosity, respectively.
The first line of Eq.~(\ref{ns0}) corresponds to the standard
Navier-Stokes equation. The second line contains spurious terms.
Nonetheless, these terms, being gradient terms with higher order derivatives, 
are negligible compared to the ones in the first
line when all hydrodynamic fields vary smoothly on the lattice scale \cite{Swift,catesjfm}.
Moreover, a convection-diffusion equation is also recovered:
\begin{eqnarray}
\partial_t \varphi + \partial_{\alpha} (\varphi v_{\alpha}) &=&
\Theta \nabla^2 \Delta \mu + \nonumber \\
&&-(\tau_{\varphi} -\frac{1}{2}) \Delta t
\partial_{\alpha} \Big(
\frac{\varphi}{n} \partial_{\alpha} p
+\frac{\varphi}{n} \partial_{\beta} c^2 P_{\alpha\beta}^{chem}
\Big) + o(\Delta t^2)
\label{conv0}
\end{eqnarray}
where the macroscopic mobility is $\Theta=\Gamma (\tau_{\varphi} -1/2)
c^2 \Delta t$.
As before, the spurious terms in the second line of Eq.~(\ref{conv0}) 
are also expected to be small.

If the density fluctuations are negligible, the continuity and 
Navier-Stokes equations (\ref{cont0})-(\ref{ns0})  
describe the behavior of an incompressible fluid. 
However, this is not always the case in LBM where
the compressibility effects can produce errors in numerical 
simulations.
In the next two subsections we will modify the described LBM in order to
reduce unphysical compressibility effects.  

\subsection{The limit of small Mach number}

The first scheme that we consider and adapt to our LBM is the one introduced 
in Ref.~\cite{heluo}.
Previous studies for incompressible fluids have shown that 
the density fluctuations, $\delta n$, are expected to be of order $o(M^2)$ in the limit
$M \rightarrow 0$, $M=v/c_s$ being the Mach number \cite{zou95,martinez}.
We enforce this condition in the method.
By substituting $n=n_0+\delta n$ into the equilibrium distribution functions (\ref{svil1}), 
$n_0$ being the specified constant value of density,
and neglecting the terms proportional to $\delta n ({\bf v}/c)$,
and $\delta n ({\bf v}/c)^2$, which are of the order $o(M^3)$ or higher, 
we find that the 
net effect is to replace $n$ with $n_0$ into
the expansion coefficients (\ref{bs})-(\ref{ds}) keeping unchanged the rest
of the scheme.
In this way the terms in the equilibrium distribution functions (\ref{svil1}) 
are of the order $o(M^2)$ or lower. The condition
$M \ll 1$ is satisfied in our runs where it is always $M < 0.01$.

In this case the Chapman-Enskog expansion gives 
the continuity equation
\begin{equation}
\partial_{\alpha} v_{\alpha} = 0 + o(\Delta t^2) + o(M^2),
\label{cont1}
\end{equation}
the incompressible Navier-Stokes equation
\begin{equation}
\partial_t v_{\beta} + v_{\alpha} \partial_{\alpha} v_{\beta} =
-\frac{1}{n_0} \partial_{\beta} p  
-\frac{1}{n_0} \partial_{\alpha} c^2 P_{\alpha\beta}^{chem} +
\nu \nabla^2 v_{\beta} + o(\Delta t^2)+o(M^3),
\label{ns1}
\end{equation}
and the convection-diffusion equation
\begin{equation}
\partial_t \varphi + \partial_{\alpha} (\varphi v_{\alpha}) =
\Theta \nabla^2 \Delta \mu + o(\Delta t^2)+ o(M^3),
\label{conv1}
\end{equation}
where we have not written again the spurious terms of Eqs.~(\ref{ns0})-(\ref{conv0}).

\subsection{A feedback mechanism}

In Ref.~\cite{fang} a different algorithm based on a feedback 
mechanism is proposed to keep the density nearly 
constant in LBM for a single fluid. 
Here we adapt this scheme to our model combining it with 
the one previously described.
We use for the expansion
coefficients (\ref{as}) the following form:
\begin{eqnarray}
A_0 &=& a_0 n - \frac{5}{6} P_{\alpha\beta}^{chem} \delta_{\alpha\beta} 
\nonumber \\
A_I &=& a_I n + \frac{1}{6} P_{\alpha\beta}^{chem} \delta_{\alpha\beta}
\label{newa0} \\
A_{II} &=& \frac{A_I}{4} \nonumber
\end{eqnarray}
Let us note that these expressions coincide with (\ref{as}) when $a_0=4/9$ 
and 
$a_I=1/9$. Here we only ask that $a_0+5 a_I =1$ to guarantee the total density 
conservation. 
We consider at each node and at each time step before collision 
the density $n$ fixing the value
of $a_I$ as
\begin{equation}
a_I=a_I^0 - b (1-\frac{n}{n_0})
\label{ai}
\end{equation}
where $a_I^0=1/9$ and $b$ is a positive constant given by experience.
In this paper we generally use $b=0.01$. The value of $b$ is flexible, 
but if it is too large the simulation will be unstable.
The value of $a_0$ follows from the relation between $a_0$ and $a_I$.
These values are then used in (\ref{newa0}).
This choice forces, in the next streaming step, more (less) particles
to leave the node with $n > n_0$ ($n < n_0$), resulting in 
$n \rightarrow n_0$.

By using a Chapman-Enskog expansion and neglecting terms of order 
$o(M^3)$ or higher we find 
the continuity equation (\ref{cont1}),
the incompressible Navier-Stokes equation (\ref{ns1})
where now the isotropic pressure contribution is 
\begin{equation}
p^{'}=3 a_I c^2 n = c_s^2 n - 3 c^2 b (n_0 - n) + o(\delta n^2) 
\label{presscal2}
\end{equation}
and the convection-diffusion equation (\ref{conv1}).
For the convenience of description, we label the old scheme as A, the scheme
described in subsection B as B, the scheme combining scheme B and the feedback
mechanism as C. 

\section{Phase separation with periodic boundary conditions}

After a binary mixture is quenched into the coexistence 
region and  domains of the two phases are well established,
the typical size of domains generally grows as
$R(t) \sim t^{\alpha}$. 
A simple scaling analysis of the Navier-Stokes and 
of the convection-diffusion 
equations shows that three regimes can be found
depending on  the role played by  hydrodynamic degrees of freedom.
At high viscosity the domain growth is governed 
by a diffusive mechanism and  
$\alpha = 1/3$ \cite{Lifshitz61}.
When  hydrodynamics becomes relevant,
the  laws $R(t) \sim t$ or  $R(t) \sim t^{2/3}$
are expected depending on whether viscous forces or inertial 
effects dominate, respectively \cite{Furukawa85,Bray94}. 
However, in real systems the situation
is more complex. 
The physical mechanism responsible for  viscous growth
is not operating in the two-dimensional case \cite{Siggia79}
and, indeed, this regime
has never been observed in simulations \cite{Yeomans}.
In this paper, since we address the question of stability
which becomes  more critical at low viscosities,
we will focus on cases corresponding to the inertial regime.

The average size of domains $R(t)$ can
be calculated as 
the first momentum of the structure factor, that is 
\begin{equation}
R(t) = \frac{ \int d \vec k \;\;C(\vec k,t)}{\int d \vec k \;\; |k|
\;\;C(\vec k,t)}
\end{equation}
where 
\begin{equation}
C(\vec k,t)= \langle \varphi(\vec k, t)\varphi(-\vec k, t)\rangle   \qquad 
\end{equation}
and $ \langle \cdot \rangle $ is the average over different histories.

In Fig.~1 we show the behavior of $R(t)$ for the sets of parameters
listed in Table I. Similar behaviors have been also observed in many other cases 
here not reported.
We have also not reported the 
 worst situations occurring  at very low $\tau$, when
the scheme A gets very soon unstable ($t \simeq 100$) and  
the new schemes are not significantly better.
All the cases of Table I refer to situations where phase separation
has already generated well-formed macroscopic domains
and a significant evolution of $R(t)$ can be observed.
For each set, results obtained with the schemes A,
B and C are compared.
While  the scheme A can be used only before
density fluctuations start to grow indefinitely,
the simulations with the new schemes remain stable at all times for most of
the cases considered.
We have found only few cases (e.g., set 6 at the lowest
considered viscosity in Table I)
where the schemes B and C
are affected by instability; however, the instability
with the scheme C occurs when the system is almost completely phase separated,
later than with the scheme B and
much later than with the scheme A.

All the cases shown correspond to the inertial regime and for
reference we have plotted the line with slope $2/3$. 
The sets 1 and 2 only differ in the value of the mobility $\Gamma$.
The main difference is that, before reaching the hydrodynamic $2/3$ regime,
the growth is faster when  the  mobility is higher
as it can be observed in Fig.~1. 
The slowing down of growth at large times is due to
finite size effects appearing when the size of domains 
becomes comparable with the lattice size.

In order to understand the effects of density fluctuations,
we show in Fig.~2  pictures of the system 
for the parameter set  2 at the time $t=5895$ when 
 the scheme A becomes unstable.
Simulations  obtained with the old and the new schemes can be compared
using  the same initial conditions.
Differences in the order parameter 
field can be observed due to the different treatment
of density fluctuations. When the scheme A is used
the density plot shows the appearing of
a peak which will grow indefinitely.
The presence of the instability can be also observed in the 
middle low region of the concentration field as a white stripe. 
In the case with the 
scheme B fluctuations remain  limited around the average value $n=1$
and are larger  in correspondence of interfaces. The velocity field also
shows an instability with the scheme A while with the new scheme
regular local flows can be observed.

>From sets 3 and 6 of Fig.~1 we observe that $R(t)$  behaves 
more regularly when calculated with the scheme C. 
This allows 
a better determination of the growth exponent.
The effects of
using a bigger lattice size can be seen
in Fig.~1 for cases 4 and 5.  

In recent simulations of three-dimensional 
systems based on the original algorithm the growth exponent
was obtained by fitting the results of $R(t)$ with the function 
$f(t) = c(t-d)^{\alpha}$ in a time interval from
 the inflexion point to the onset of the instability \cite{catesjfm}.
This procedure, as admitted by the same authors of Ref.~\cite{catesjfm} and
discussed also in Ref.~\cite{coveney},
is very sensitive to small variations 
of the initial time considered, so that the evaluation 
of the growth exponent is problematic.
We have applied a similar procedure to our results
confirming the sensitivity to small variations 
of the initial time considered and finding different values for the exponent
when the old or the new more stable schemes are applied.
For example, for the set 2 with the old method,
we find $\alpha = 0.58$ if we start from the inflexion point located at $t=3000$
and $\alpha = 0.61$ if data are fitted from $t = 4000$ (the run becomes 
unstable at t = 5985). If we use the longer series of 
data obtained with the scheme C
we get $\alpha = 0.64$ starting from the inflexion point or at a later time.
For the set 4 the fitting procedure for the scheme A
gives values of $\alpha $ in the range $0.50 - 0.55$ 
changing the initial time for the fitting procedure, 
while the scheme B
gives values in the range $\alpha = 0.64 - 0.67$.
Other cases are also worst because sometimes, like for the set 3, 
a reliable determination of the exponent is not possible with the old scheme.

\section{Stability with walls}

In this Section we discuss how the presence of boundary conditions
with walls affects the stability of simulations of phase separation.
Boundary conditions with walls are present  in problems where 
a flow is applied to the system, e. g. shear flow,
or in situations with geometrically complex boundaries.

Fluctuations of density are generally larger close to the walls
due to the fact that the propagation and collision steps involve
a reflection on the boundary.
Then, if the density is larger than the average value in a point close 
to a wall,
the reflection tends to enhance this fluctuation.
An example of the behavior of density just before the 
system becomes unstable is given in Fig.~3 which shows clearly
that the instability appears close to the walls.

Also for the case with walls we have  applied
the schemes described in Sec. II for controlling
the density fluctuations. 
In all the considered cases  instabilities in
simulations with walls occur  much before 
than in simulations with periodic boundary conditions.
In Table II we have reported the stability results for the 
original and the new scheme C. 
The comparison shows that in most of the cases
the new algorithm can be very useful for running longer
simulations of phase separation.

\section{Conclusion}

In this paper we have considered two-dimensional 
Lattice Boltzmann methods for incompressible
binary mixtures 
based on a  free-energy functional approach.
We have introduced mechanisms for controlling the total density
fluctuations. These mechanisms were already shown useful 
for reproducing more accurately examples of analytical solutions 
for simple fluid flows.

In the case of binary mixtures,  density fluctuations
can induce numerical instabilities that  
make impossible to  run long enough simulations as it is needed
in problems like spinodal decomposition.
We have used the new algorithms 
to study the phase separation of binary mixtures
in the inertial regime at low viscosities when the effects of
density fluctuations are worst.
We have found  that the stability of the simulations 
can be largely improved. For  most of the parameters considered,
differently from what happens with the original algorithm,
it is possible to take under control density fluctuations eliminating
the occurring of instabilities. This allows to study  the properties
of phase separating binary mixtures more reliably; in particular,
a more accurate evaluation of the growth exponent
can be given. 
As a future research in this direction, it would be interesting to 
apply  the  LBM with improved stability to
 the case of a three-dimensional mixture.

\acknowledgments
A.L. acknowledges INFM for partial support.

\newpage

\begin{table}
\begin{center}
\centerline{\epsfig{file=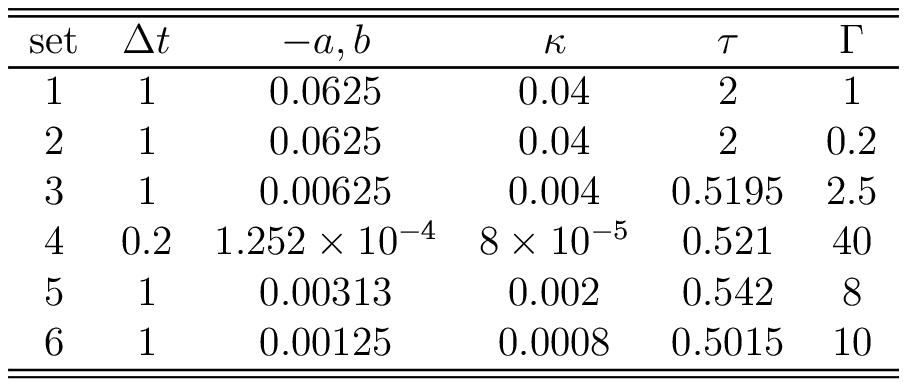,bbllx=184pt,bblly=599pt,bburx=447pt,bbury=710pt,
width=0.8\textwidth,clip=}}
\end{center}
\caption{The sets of parameter used in Figure 1.}
\label{table1}
\end{table}

\newpage

\begin{table}
\begin{center}
\centerline{\epsfig{file=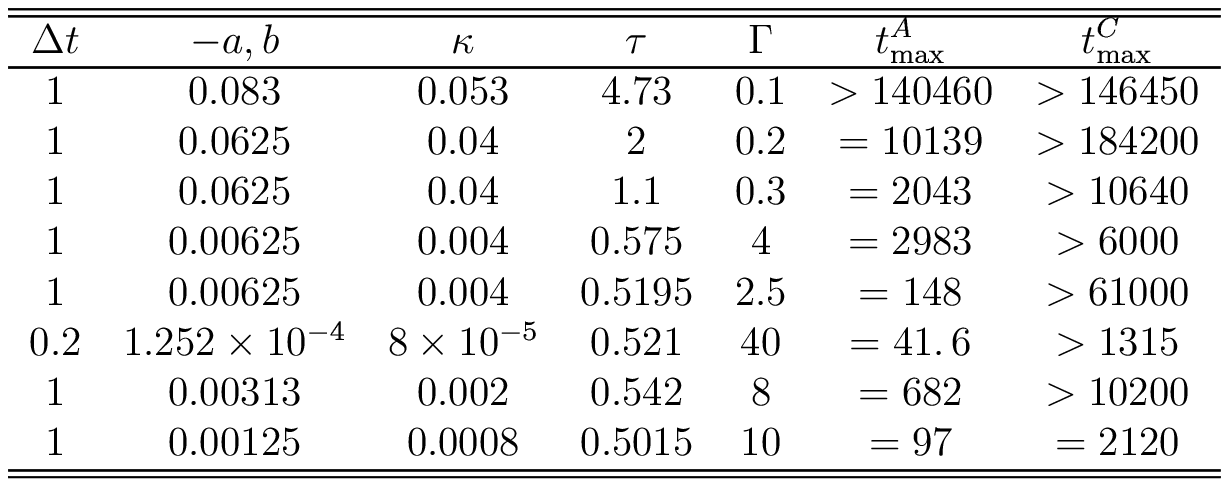,bbllx=139pt,bblly=568pt,bburx=494pt,bbury=711pt,
width=0.8\textwidth,clip=}}
\end{center}
\caption{Safe running times for the schemes A and C for 8 
sets of parameters. $t_{max}$, where the upper label refers to the scheme used, is the longest simulated time.
The symbol $=$ means that the simulation becomes unstable after that time; the symbol $>$ means
that the simulation is still stable up to that time.}
\label{table2}
\end{table}

\newpage

\begin{figure}
\begin{center}
\centerline{\epsfig{file=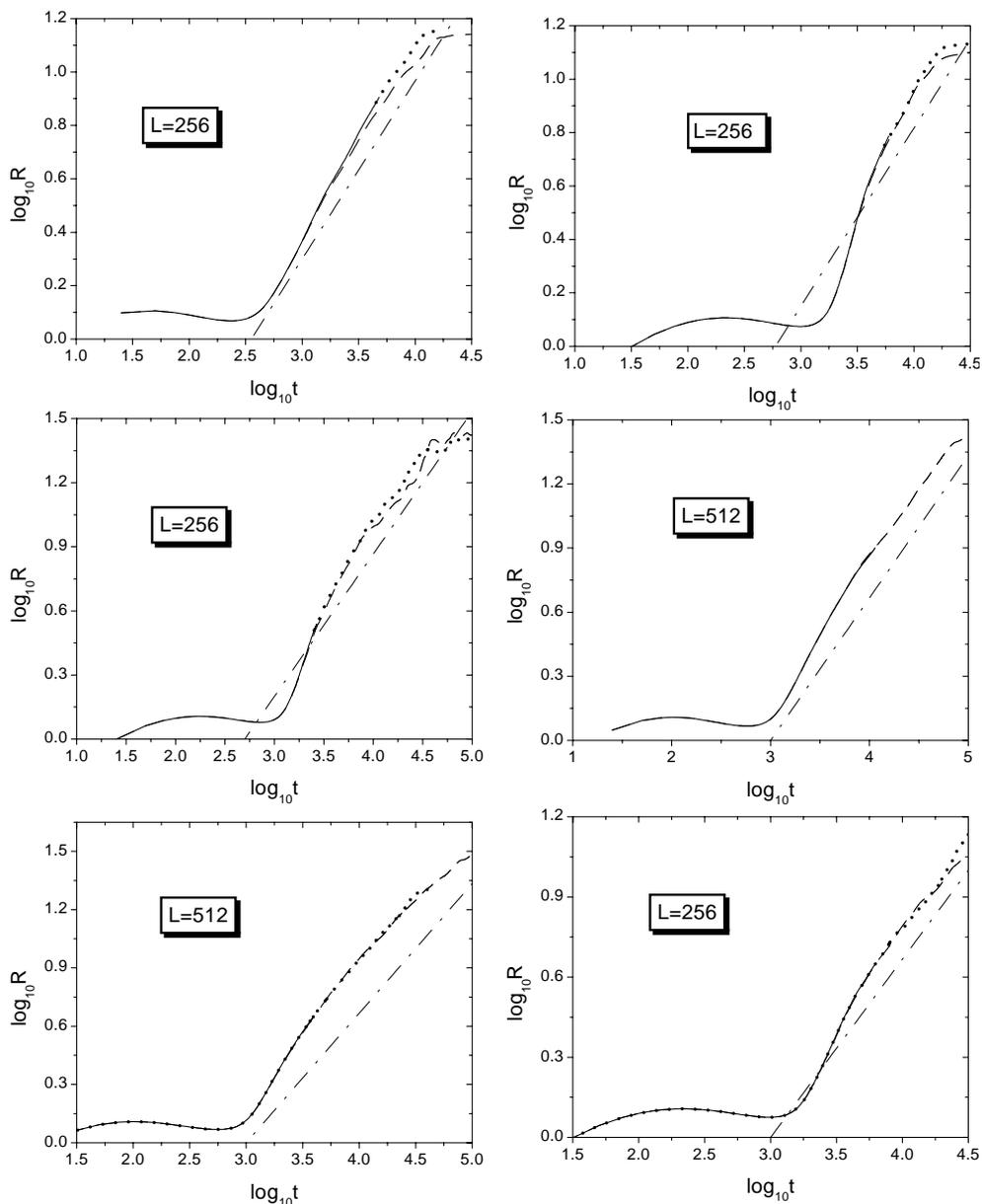,bbllx=70pt,bblly=192pt,bburx=527pt,bbury=754pt,
width=0.8\textwidth,clip=}}
\caption{$R(t)$ behavior for the sets of parameters listed in Table I. The solid, dashed, and dotted lines
correspond to scheme A, B, and C, respectively.
Runs with the schemes A and B always start from random initial configurations. The scheme C is applied either
to simulations from random initial conditions (sets 5 and 6) or starting from a configuration slightly before 
the scheme A fails to work. Runs with the scheme C were eventually stopped, though still stable, once the system
was phase separated.}
\end{center}
\label{fig_1}
\end{figure}

\newpage

%\begin{figure}
\begin{center}
Fig.2 and Fig3 are in JPG format. Captions of Fig.2 and Fig. 3 are as follows,
\end{center}
%\centerline{\epsfig{file=Fig2.eps,bbllx=106pt,bblly=155pt,bburx=410pt,bbury=710pt,
%width=0.6\textwidth,clip=}}
%\caption{

Fig. 2 Configurations of the order parameter $\varphi$ (a)-(b), the total density $n$ (c)-(d), 
and velocity fields (e)-(f) for the parameter set 2 at the time
$t=5895$. Figures (a), (c), (e) are for the scheme A, while (b), (d), (f) are for the scheme B. 
In picture (a) it is $-2.13 \le \varphi \le 1.77$ and in (b) it is $-1.06 \le \varphi \le 1.05$.
%}

%\end{center}
%\label{fig_2}
%\end{figure}

%\newpage

%\begin{figure}
%\begin{center}
%\centerline{\epsfig{file=Fig3.eps,bbllx=109pt,bblly=375pt,bburx=493pt,bbury=819pt,
%width=0.6\textwidth,clip=}}
%\caption{

Fig.3 Three-dimensional plot of the total density $n$ in a case with walls 
at time $t=1315$ using the scheme B.
The parameters are those of set 4 in Table I.
%}
%\end{center}
%\label{fig_3}
%\end{figure}

\end{document}